# Interactive Data Integration through Smart Copy & Paste

Zachary G. Ives[1]    Craig A. Knoblock[2]    Steven Minton[3]
Marie Jacob[1]    Partha Pratim Talukdar[1]    Rattapoom Tuchinda[4]
Jose Luis Ambite[2]    Maria Muslea[2]    Cenk Gazen[3]

[1] Univ. of Pennsylvania, {zives,majacob,partha}@cis.upenn.edu

[2] Univ. of Southern California, {knoblock,ambite,mariam}@isi.edu

[3] Fetch Technologies, {sminton,gazen}@fetch.com

[4] National Electronics and Computer Technology Center (Thailand), tuaguan@gmail.com

## ABSTRACT

In many scenarios, such as emergency response or ad hoc collaboration, it is critical to reduce the overhead in integrating data. Here, the goal is often to rapidly integrate "enough" data to answer a specific question. Ideally, one could perform the entire process *interactively* under one unified interface: defining extractors and wrappers for sources, creating a mediated schema, and adding schema mappings — while seeing how these impact the integrated view of the data, and refining the design accordingly.

We propose a novel *smart copy and paste* (SCP) model and architecture for seamlessly combining the design-time and run-time aspects of data integration, and we describe an initial prototype, the CopyCat system. In CopyCat, the user does not need special tools for the different stages of integration: instead, the system *watches* as the user copies data from applications (including the Web browser) and pastes them into CopyCat's spreadsheet-like workspace. CopyCat *generalizes* these actions and presents proposed auto-completions, each with an *explanation* in the form of provenance. The user provides *feedback* on these suggestions — through either direct interactions or further copy-and-paste operations — and the system learns from this feedback. This paper provides an overview of our prototype system, and identifies key research challenges in achieving SCP in its full generality.

## 1. INTRODUCTION

Today's data integration tools require substantial up-front investment at *design-time* — understanding source schemas, creating a mediated schema, defining schema mappings — before a *runtime* system can be used to produce results or handle updates. Thus integrating data is a long and laborious process: by the time good results are achieved, the "window of opportunity" where the data is most useful may have elapsed, or application requirements might have changed!

This has led to a series of research efforts designed to reduce the initial design-time effort (perhaps sacrificing some result quality): *dataspaces* [11], "pay as you go" data integration [31], "best effort" integration and extraction [27, 32], and peer-to-peer query answering with *composable* [14, 17] or *probabilistic* [10] schema mappings. Such work shares two tenets: (1) provide basic functionality even when little user effort has been invested, and more functionality as more human input is given; (2) leverage and reuse human effort where possible. The expectation here is that the integration process will be done by iteratively switching between design-time and runtime, until sufficient result quality is achieved.

In this paper we argue that, for many "best effort" integration applications where time is of the essence, an even better approach is to *combine* design-time and runtime aspects into a single *interactive* process. One should be able to add sources, design a target schema or even a one-off query, specify mappings or other operators, see results, and refine those results or the schema — all on-the-fly, with a single seamless mode of interaction. This enables a data integrator to develop an *understanding* of the data sources as he or she is integrating them; and to assemble and revise the integrated or mediated schema and the mappings in accordance with this understanding. Importantly, the integrator can directly *see* the impact of design choices on the integrated data (which is also "explained" by visualizing its *provenance* [4, 8, 17]) as they work. Such a model is especially appropriate for "one-time" data integration tasks, where the goal is to integrate results to answer a specific query.

To support this type of integration, we propose and implement a scheme we term *smart copy and paste* (SCP): the integrator follows the familiar model of copying items of interest from existing applications (Web browser, office applications, etc.), and pasting them into a dynamic, spreadsheet-like "workspace."[1] As the integrator pastes content, the system attempts to infer potential *generalizations* of the user actions, and it shows *suggestions* along the lines of an "auto-complete" in Microsoft Word. These are intuitively the results from proposed *information extractors* (wrappers) over data sources, and from potential *mappings* (transformations expressed as queries or constraints) across sources.[2] The integrator provides *feedback* to the system by accepting the auto-complete suggestions (or portions thereof), or by simply ignoring the suggestions and pasting further content. Finally, an SCP system should include built-in interfaces to data visualization tools such as Google Maps, as well as the ability to export data to standard formats. To illustrate how SCP works, we sketch a basic usage scenario below for our prototype system, CopyCat (**Copy** and con**Cat**enate).

EXAMPLE 1. *Consider a hurricane relief scenario: FEMA needs to establish connections to supplies, shelters, road conditions, dam-*

---

[1] The appearance of this workspace could be configured differently, but we feel that the spreadsheet metaphor is the most general one for table construction.

[2] SCP can be used either to define integration queries or to construct mappings forming integrated views of data.

*age regions, etc., and begin making decisions. This is a* best effort data integration *problem: it is more critical to immediately get data (even with a few errors) than it is to get perfect results.*

*Suppose one integration task is to take a list of shelters from a television news Web site, combine it with the shelters' contact information from a spreadsheet, and plot the shelters on a map. In CopyCat, a data integrator would load the page of shelters into her Web browser. She would select and copy the first item, then paste it into the CopyCat workspace. The system would try to* generalize *the integrator's action by extracting* other shelters *from the same page and proposing new rows on the workspace. She might accept these new rows and then copy the first shelter's name into Google Maps to get its full address and geocode. She would paste the resulting information into the workspace, in the same row as the first shelter. The system would again generalize, taking all subsequent shelter names, feeding them to the map site, and retrieving the matching addresses and geocodes. In some cases the shelter name may be ambiguous and might return multiple answers: here CopyCat would show the alternatives and allow the integrator to select the appropriate location.*

*Finally, the integrator would load the spreadsheet of contacts, and copy the contact info that best matches the first shelter (e.g., approximately matching the name and address), then paste it into the workspace. Here the match might not be a direct lookup, but rather the result of approximate* record linking *techniques, which determine the contact that best matches each shelter. CopyCat* learns *the best combination of heuristics for this case of record linking, via a combination of generalizing examples (the integrator might paste matches for several shelters) and accepting feedback (she might accept or reject suggested matches).*

*Ultimately, a complete table would be assembled in the workspace. This might itself represent a "one-off" data integration query focused on answering a single question. Alternatively, it could be persistently saved as an integrated, mediated view of the data, enabling user or application queries over a unified representation. In our example, the data would also be* exported *to a Google Maps visualization.*                                                                 □

Note that interaction with an SCP system is quite different from running the usual series of schema matching, mapping creation, and query processing tools. In our model, mappings are defined by copying data from existing applications to the clipboard, and then pasting to the SCP workspace. The SCP system's semi-automatic features suggest auto-completions that can be ignored at no cost, or accepted or refined if beneficial. The integrator gets to see and provide feedback on *data* the moment she provides input to the system. We describe the user interface in detail in the next section.

We note that our initial work on SCP focuses on a specific class of integration tasks. The data integration problem comes in many forms, and no single solution addresses all requirements. For instance, enterprise-wide information integration typically involves gathering large volumes of data from tightly controlled sources, in order to perform OLAP queries or other analysis. Another common scenario is that business or scientific data is shared among a small number of parties, and portals or applications are developed to support a limited number of predetermined queries. Our focus in this project is instead on data integration problems where a moderate number of Web and document sources (each with KB or MB of data, but probably not GB) need to be integrated in a *time-sensitive* manner, possibly with small databases. In our target settings, a set of sources might be integrated *on-demand to answer a specific query* or class of queries; or they might be integrated in a way that rapidly *evolves the mediated schema* as new data is incorporated and new questions are posed.

In the remainder of this paper, we describe our initial design and implementation of our CopyCat prototype, our experiences with it, and an agenda for future research. In Section 2, we illustrate the SCP user interface and show its architecture, and then outline how our CopyCat system implements these functionalities. In Section 3, we describe how CopyCat *learns* the structure of documents and the data types they are manipulating, such that we can determine what is being integrated. Then in Section 4, we describe how one can build complex schema mappings or queries using our tool. We next describe open research challenges in Section 5. Finally, we discuss related work in Section 6 and conclude in Section 7.

## 2. SMART COPY AND PASTE

Smart copy and paste is a form of *programming by demonstration* [9, 25, 36] (PBD): users *demonstrate* the actions to be performed to integrate data (copying data from source applications to the SCP workspace). The system *learns to generalize* their actions. Then the system immediately shows the effects of applying these generalizations, in the form of auto-complete suggestions, and solicits *feedback* on these suggestions. Through *provenance* information, the feedback can be "traced back" from data to the transformations (mappings) and extractors that are responsible for the data.

In this section we describe the SCP user interface, introduce the basic architecture of a generic SCP system, and finally outline how our CopyCat implementation works. In subsequent sections we describe the operation of the main modules in our system.

### 2.1 User Interaction with SCP

To explain how a user interacts with an SCP system, we refer to a pair of screenshots from our CopyCat implementation.

**Adding a data source.** Initially, as the user pastes data from one source, the SCP system is in *import mode*. In Figure 1 the user has pasted into the table at the top of the window (the CopyCat Workspace) two entries from the list of shelters from Example 1. Based on data patterns seen previously, the SCP system determines that the second and third columns represent street addresses and cities, and suggests the column types PR-Street and PR-City with names Street and City. The user manually enters the label for (shelter) Name. Additionally, the system takes the pasted rows and suggests *row auto-completions* (highlighted in the figure), by finding an information extraction pattern that returns the first two shelters plus additional ones. The user may provide feedback on these suggested rows, specifying that they should be kept or removed. This feedback gets sent to the source learners, which will refine the extraction pattern, e.g., to include or exclude certain HTML tags, data values or document delimiters in its matches.

**Integrating data.** The user can switch the SCP system into *integration mode* by clicking on a button, or by pasting data from a different source into a contiguous row or column in the current workspace (such a paste expresses a union or join). In this mode, the SCP system will create a tabbed pane in its GUI for each data source. The user may select any of these, and then either paste new data or accept the SCP system's suggestions for new rows or columns. The moment the user pastes or accepts a row or column from a different source into the current tab, the system recognizes that a query has been constructed, and the query's output receives its own tabbed pane.

If the query is being created by pasting data, then the SCP system may spawn off a background task to import the source of that pasted data, much as described above. Then it must identify which query the user has been trying to construct by pasting data from two sources into the same table: this involves discovering the join conditions or record linking operations necessary to associate the columns. The SCP system will next auto-complete the new columns' values for each row (this represents a **row autocompletion**).

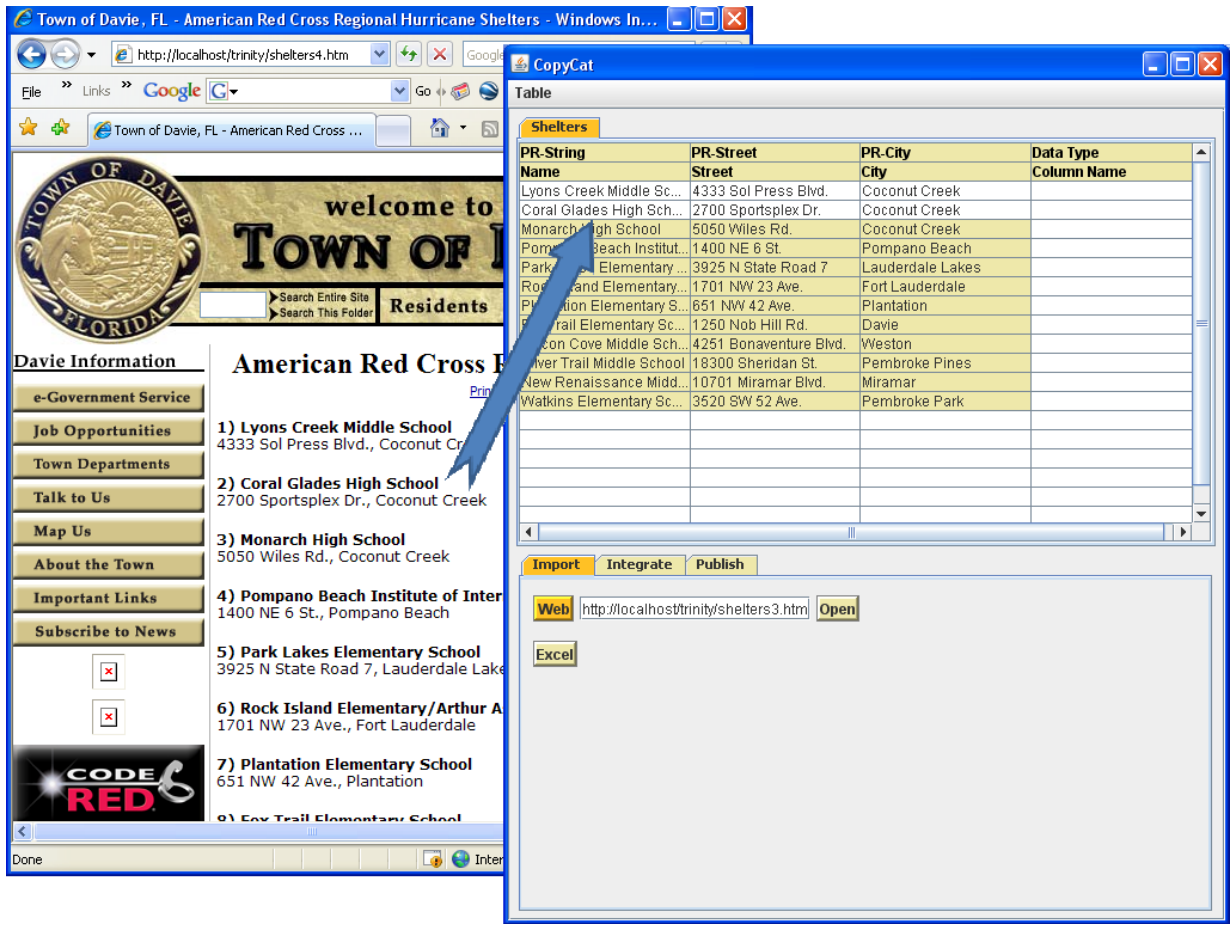

Figure 1: Screenshot of CopyCat's import mode as shelters are being added

Alternatively, as in the screenshot from CopyCat in Figure 2, the SCP system may already have imported a relevant source and it may suggest one or more **column autocompletions** to the user. In our example, CopyCat has existing knowledge of several data sources and Web services, including a zip code resolver that uses Google Maps to find zip codes using address information. It thus suggests a Zip column (rightmost attribute, highlighted in yellow) as the most promising auto-completion, possibly from among several alternatives.

**Explanations and feedback.** In some cases, there may be erroneous tuples in the suggestions. An SCP system emphasizes helping the user *understand* the presence of a suggested tuple or set of attributes (in a similar vein to debugging schema mappings [6]), so he or she can provide *feedback* on what has been integrated. The Tuple Explanation pane (bottom of Figure 2) visualizes the provenance of the selected tuple in the table. Three attributes originate from the Web page called Shelters. The Street and City values are fed into the Zipcode Resolver (a *dependent join*; illustrated by the directed arrows to attributes in the rightmost table, and the green color scheme), which yields a Zip attribute. From the pop-up context menu in this mode, the user may explicitly accept or reject the suggested column auto-completion. This feedback operation is fed to learning components in the SCP system, as discussed in the next section. The system will revise its auto-complete suggestions accordingly.

## 2.2 Generic SCP Architecture

Figure 3 illustrates the architecture of an SCP system. Copy and paste operations — between source applications and the SCP workspace — are detected by application **wrappers**. Monitored operations, as well as context information like the document being displayed in the source application, are fed into three *learner* modules.

Two learners focus on properties of individual sources: the **structure learner** learns extractors that crawl the document structure of the source (including hierarchical Web sites as well as documents or forms with multiple segments), and it discovers any necessary input bindings required by the source. The **model learner** determines attributes and a schema for the source. The resulting source description gets added to a system catalog.

The **integration learner** determines the *query* and/or set of *mappings* that the user is creating by copy-and-paste. If the user performs a series of copy-paste-search-copy-paste operations, perhaps each intervening search represents a dependent join: attributes from the first source are looked up in the second source, and so on. In some other cases the user wants to perform a *record linking* or *approximate join* operation: here the SCP system can attempt to learn a record linking function from a set of examples — or, in some cases, use a function from a predefined library.

Once a few examples have been provided, the learners attempt to generalize from them, finding potential extractors and transformations. A ranked set of promising extractors and queries is produced by the **auto-complete generator**.

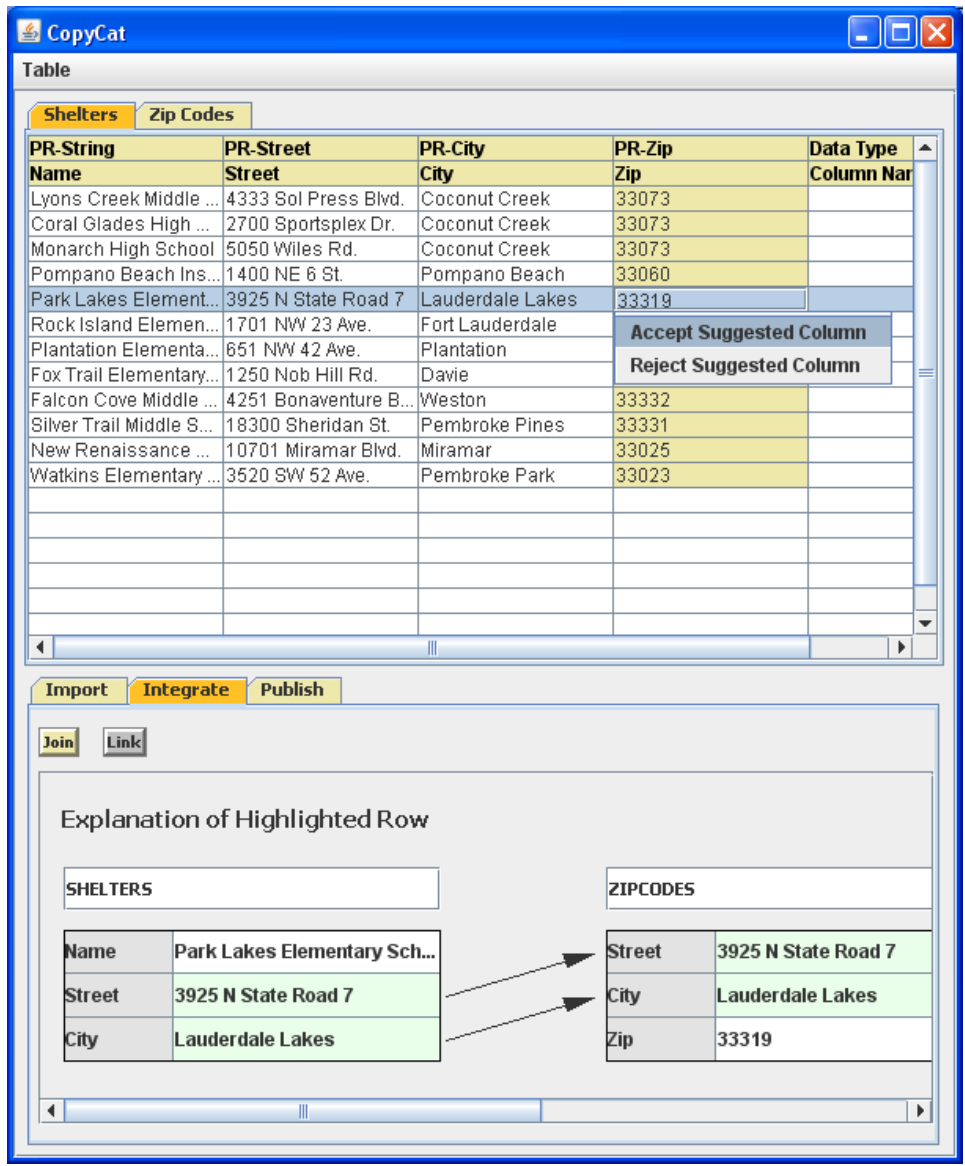

Figure 2: Screenshot of CopyCat's integration mode with a suggested column auto-completion

In turn these queries are run by the **query engine** to produce example answers, which are output to the user as extra rows and columns in the **workspace**. The user may provide *feedback*: promoting or demoting tuples, modifying the headings or data type specifiers for the columns, or adding or removing columns. Each of these actions provides information to the learners in the system. Through *data provenance* [4, 8, 17], feedback on tuples can be related back to the tuples' source queries. The learners adjust source scores, extraction patterns, and record linking or join conditions, in order to respond to the user feedback.

### 2.3 CopyCat: A Prototype SCP System

In the remainder of this paper, we describe the details of our initial CopyCat prototype, which builds upon the authors' prior experience with learning-based data integration tools. Our focus in this prototype is on the coupling between the clipboard, the workspace/user interface, and the learning systems.

**Application wrappers.** The initial CopyCat prototype supports monitoring of copy operations from a variety of common applications: Web browsers like Internet Explorer, and any HTML forms or pages they display; and Microsoft Office applications like Word and Excel.

**Structure learner.** CopyCat is given direct access to the underlying data being displayed in the browser or application window, as well as the data being copied. Our structure learner (Section 3.1) seeks to identify the origin of the copied data and to generalize the extraction operation(s) being performed, across the structures of the source data and — for multi-page sources — the source hierarchy.

**Model learner.** This component (Section 3.2) detects the types of the data items being manipulated, and ultimately a source model that describes the function performed by the source. It uses this information to help find possible *associations* among data items it knows, or operations that can be performed to find related data: for instance, a join might be used across sources on social security number; a phone number might be looked up in a reverse directory to find a person; a record-linking function might match addresses.

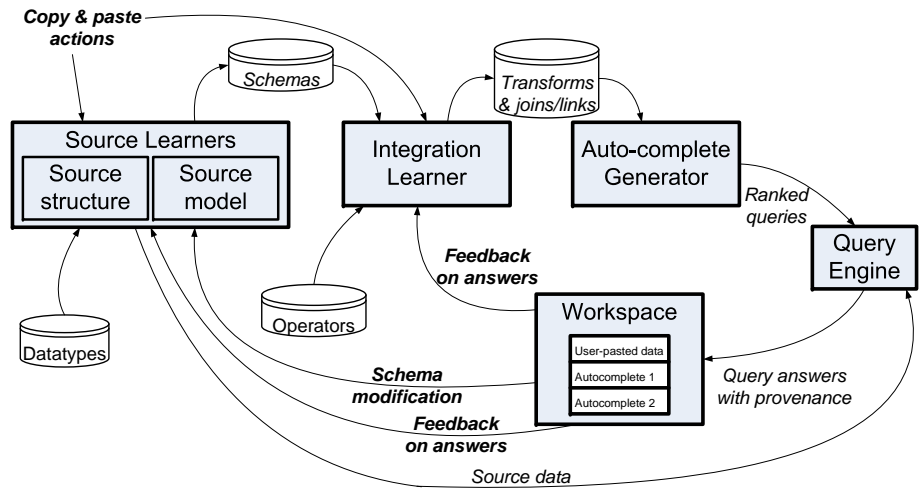

Figure 3: Architecture diagram for SCP system, with user interactions highlighted in boldface

**Integration learner.** Given a set of possible associations with different scores, the integration learner (Section 4.2) determines the most promising queries, which define auto-completions. It processes user feedback over these auto-completions to re-weight the scores of different associations, thus learning the user's preferred set of associations and integration queries.

**Query engine.** CopyCat employs the ORCHESTRA query answering system [17], which builds a layer over a relational DBMS to annotate every answer with data provenance. As described previously, provenance enables CopyCat to convert feedback on auto-complete data into feedback over the *queries* that created the data.

**Workspace.** The initial CopyCat interface is implemented in Java Swing, and is shown in Figures 1 and 2. While it provides both row- and column-auto-completions, its emphasis is on suggesting columns one at a time.

In the remainder of this paper, we describe how the three learning components interoperate with the workspace. We refer the reader to [34] for an overview of how data provenance is recorded and maintained by the query processor.

## 3. LEARNING ABOUT SOURCES

Logically, each source has associated with it an extractor and a schema (possibly including input binding restrictions). The source structure learner is primarily responsible for learning the extractor, and the source model learner is primarily responsible for learning the schema; however, the two modules are closely linked and share information.

### 3.1 Structure Learner

The structure learner analyzes data that is copied into the CopyCat workspace. By analyzing the data and the context from which it was copied, the structure learner can generalize the copy-and-paste operation, providing a set of auto-complete suggestions that make it simple for the user to extract and import data from a source.

For a relatively structured source such as an Excel spreadsheet, the generalization process is normally quite simple. For example, after copying just two data items from a column in spreadsheet, it is clear that the user's selection should be generalized to include all the additional rows in that column with similarly-typed information. For semi-structured sources, such as Web sites, the hypothesis space is much larger. For instance, in the example in Figure 1, after copying the data about the first two shelters, both of which are in Coconut Creek, it is not immediately clear whether the proper generalization is to copy the entire list, or copy just the shelters in Coconut Creek. The space of possible generalizations can be considerable, especially when the sources are complex. For example, CopyCat can extract data from a web site where there are multiple pages (e.g, pages accessible via a form), each of which may have complex lists of data that should be extracted into the CopyCat workspace. If these pages are well-structured, a single example can be illustrative enough that the system correctly generalizes across all the pages. However, the more complex the pages are, the more examples may be necessary for the system to induce the correct generalization.

After each copy and paste operation, the structure learner guesses a generalization, and the user can provide feedback to the system either by accepting or rejecting the auto-completed suggestions. If the user rejects the suggestions, the system will choose another hypothesis and revise the suggestions. If the user pastes another data item into a row (either replacing a current suggestion or augmenting the current suggestions) the system will select a new hypothesis and make appropriate suggestions.

This learning approach is based on our previous work on extracting data from the Web, where the learner analyzes the structure of a website to identify its relational structure [12]. In summary, the approach works as follows: First, a set of software "experts" analyze the given set of pages. Each expert is an algorithm that generates hypotheses about the structure the web site, focusing on a particular type of structure. For example, we have experts that can induce common types of grammars for web pages, experts that analyze visual layout information, experts that can parse particular data types such as dates, experts that look for patterns in URLs, and so on. These experts discover similarities between the various pieces of data on the site, and output their discoveries as hypotheses about the overall relational structure of the data on the site. Next, via a clustering approach, the algorithm produces its guess as to the best overall relational description of the data on the site. Essentially, this gives us a tabular view of the data on the site.

Finally, given one or more examples selected by the user, the system attempts to find a most-general projection hypothesis consistent with the example selected by the user. If this method cannot find a consistent hypothesis, the system falls back on a sequential covering approach based on more traditional wrapper induction techniques [28].

One advantage of analyzing the source's relational structure independently of the copy-and-paste process is that we need not assume that the desktop is completely instrumented. We only assume that a CopyCat wrapper provides the structure learner access to the source from which the data was selected (e.g., the Web site, spreadsheet, document, etc.) so that we can analyze the source. We do not need to know exactly where the data was cut-and-pasted from to find a hypothesis that is consistent with the copied data.

## 3.2 Model Learner

In order to understand what task the user is performing and to better support the user, CopyCat attempts to learn a model of each source that the user is manipulating. This component of CopyCat is called the model learner. It takes the results produced by the structure learner and generates hypotheses about the semantic types of the data organized into the columns of the table. As described in the next section, this capability is important for finding relevant associations across sources and also makes it possible to find alternative sources that perform the same or similar tasks. For example, if we recognize that a particular field is a social security number, then when we consider potential associations, we can consider joining this field with a field in another source that also contains a social security number.

The model learning in CopyCat has both a learning phase and a recognition phase. In the recognition phase, the system applies previously learned knowledge to recognize the semantic types of each of the columns of data that has been extracted by the structure learner. Based on the previously learned knowledge of the possible semantic types, the model learner produces a ranked list of hypotheses for the semantic type of each field. The model learner will propose the most likely hypothesis and the other hypotheses will be available in a drop down list in the CopyCat interface. The user can keep the proposed hypothesis if it is correct or select one of the other hypotheses from the drop down menu. If this is a new type of data that has not previously been seen by the system, the user can define this new type on the fly. The user simply decides on a unique name and replaces the proposed semantic type (if there is one) with the new one. The model learner will then use the data available in the source to learn to recognize this new type of information.

The approach to learning a semantic type is based on our previous work described in [24]. The system creates a set of patterns for each field using the training data available from the source. These patterns are constructed from a rich hypothesis language that includes using both the constants in the data fields and generalized tokens that describe the data, such as capitalized word, 3-digit number, etc. These patterns can be refined over time as additional training data becomes available. When the system matches the patterns against the data, there does not need to be a perfect match. Rather, the system evaluates whether the distribution of matched patterns is statistically similar to the matches on the training data. This provides a robust approach to recognizing semantic types from new sources of data that may not precisely match the original learned distribution of patterns. Once the system learns a new semantic type, this type will be immediately available in the same user session. Thus, the user can train the system on the first source that contains a given semantic type and then the system would recognize that type of field if it was available in another source that the user wanted to integrate with the first one.

In addition to learning the semantic types of the data that the user is manipulating, the model learner also tries to learn the task that is being performed by the various sources. In particular, we are focusing on learning sources that have a functional relationship between the inputs and outputs. This would include sources that map an address into the corresponding latitude and longitude coordinates (i.e., a geocoder) or sources that determine the zip code for an address. This capability allows the system to better understand a task being performed by a user and to propose sources that can fill in gaps for a user (e.g., if the zip code is missing for some of the fields) or even propose replacement sources if a source is down, too slow, or does not provide a complete set of results.

The model learner learns the function performed by a source by relating it to a set of known sources [5, 1]. The system describes the new source in terms of a set of known existing sources and then compares the inputs and outputs of the new source to the existing sources by executing the new source and the learned description and comparing the similarity of the results. In order to learn a description of a new source, the existing sources must have overlapping functionality, but this still allows the system to learn descriptions of new services that extend the coverage of known sources, compose known sources in novel ways, and provide alternative sources that perform the same functions either faster or with fewer constraints.

## 4. LEARNING TO INTEGRATE

The *integration learner* attempts to determine what integration query is being constructed, based on knowledge of data sources and possible means of combining data across sources. It executes the most likely queries and presents their results as auto-completions.

At its core, this learner maintains a *source graph* (see Figure 4), in which nodes describe the schemas of data sources and what we generically term *services*. Services can be modeled as relations that take input parameters (i.e., to use the normal data integration terminology, they have input binding restrictions). Predefined services include record-linking functions, address resolution, geocoding, and currency and unit conversion. We also model Web forms as services that require inputs. Edges describe possible means of linking data from one source to another, e.g., by joining or by passing parameters to a dependent source like a Web service. Edges receive *weights* defining how relevant they are to the integration operation being performed; the weights are typically pre-initialized to a default value and then adjusted through learning, as we discuss below.

### 4.1 Finding Potential Associations

The user manipulates the data across the different sources by copying and pasting information into the workspace. In order to support the user, the system searches for potential associations across the data sources and then uses these potential association to populate the fields in the table with possible auto-complete values. In the example shown in Figure 2, the user has extracted all of the components of an address except the zip code. Since CopyCat knows about a service (Zip Codes in Figure 4) that maps street and city into zip code, it would propose "Zip as one of the choices in the next empty column in the table. The system then invokes the appropriate source and attempts to look up the zip code for each of the addresses in the table, producing auto-completions. Similarly, CopyCat may also suggest fields from data sources that it holds in its own local repository: perhaps a previously added source may provide additional useful attributes, e.g., synonyms or supplementary information like damage estimates.

In general, there are many potential associations across the fields in the data, so a key question is how to decide when to add an association (edge in the source graph) between a pair of sources. The use of semantic types helps constrain the possible edges to add, by limiting fields to match over one or more semantic types. Nevertheless the space is still quite large. In the current system we add to the source graph edges representing joins based on (1) common attribute names and data types, (2) known links or foreign keys. We set the edge weights to a default value that exceeds the threshold necessary for the edge to be suggested in an auto-completion. If sets of sources have multiple attributes in common, we restrict the queries to match on all the attributes (i.e., we take the conjunction of all possible join predicates).

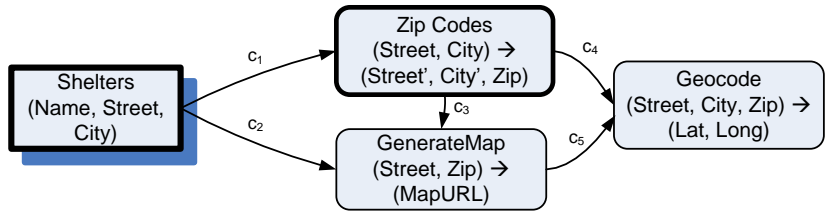

Figure 4: Subset of a source graph for our running example. Data sources are indicated with shadowed rectangles, and services with rounded rectangles. Edges represent potential associations and are annotated with *costs* $c_i$. Bolded borders indicate nodes in the query.

An important area of future work is to develop better methods that generate a relevant set of potential associations without overwhelming the user with too many choices. For instance, we would like to incorporate approximate attribute matchings, such as those from a schema matching tool [29]. Such associations are uncertain, and hence would be initialized with an edge weight that is derived from the schema matcher's confidence score.

## 4.2 Computing and Learning Top Queries

The integration learner is responsible for taking the current integration query (initially containing a single source relation) and expanding it so it produces the tuples given as examples, and possibly additional tuples. It does this by considering potential queries from a hypothesis space (the source graph), where each query receives a score that is the sum of its constituent edge weights. (This additive model has also been used in a variety of keyword search systems, such as BLINKS [20], and it enables fast computation of ranked answers as well as, in our case, efficient learning.) Given a working query, a set of tuple examples, and the source graph with weighted edges, the integration learner can determine new queries in either of two modes.

In the first case, the integration learner can provide *column completions*, as in Figure 2: it discovers promising associations (edges in the source graph scoring above a relevance threshold) from the current query's nodes to other sources. These associations may represent the joins or record linking operations necessary to "connect" these sources. For each such association, CopyCat defines a query. It creates a union of these queries (extending the schema and padding with nulls as necessary to form a homogeneous schema).

Alternatively, CopyCat may be given user-pasted tuples in which the attributes do not all originate from the same source. The tuples represent output from a join (or record linking) query, which might involve attributes and sources that have been *projected out* before the tuple was created. Here, the learner finds the most likely explanations for the tuples (queries) by discovering Steiner trees connecting the data sources in the source graph[3]. For small source graphs, we can compute the most promising queries using an exact top-$k$ Steiner tree algorithm (in our case, using an Integer Linear Programming formulation). For larger graphs we use the SPCSH Steiner tree approximation algorithm, which prunes "non-promising" edges from the source graph for better scaling. See [34] for details on the Steiner tree algorithms used and their scale-up. We note that in CopyCat, the number of sources is often relatively small, since integration (and addition of sources) is frequently done on a query-driven basis.

---
[3]A Steiner tree is similar to a minimum spanning tree, except that it starts with a set of target nodes and may add any number of intermediate nodes necessary to connect the target nodes.

Given the top queries from either algorithm, CopyCat computes results for each of the proposed queries, and gives each result the same score as the originating query. A subset of these answers are displayed to the user as auto-complete suggestions. Now the user may provide feedback to the system: either by accepting or rejecting auto-completions, or by ignoring them and pasting further data. Feedback is converted into a set of constraints. If the user accepts a group of auto-completions, they should be given a higher ranking than all alternative auto-completions; if the user rejects a group of auto-completions, these should be given a rank below the relevance threshold.

CopyCat's transformation and integration learner takes the feedback constraints and changes the weights on the source graph edges, which in turn will change the queries' relative rankings. To accomplish this, it uses a machine learning algorithm called MIRA [7, 34]. MIRA is designed for settings in which cost is computed by summing the product of *features* (in our case features are simply the nodes connected by an edge) with their independent weights. This cost model coincides with the one we use for computing costs and Steiner trees. In order to learn new weights, MIRA first compares the nodes and edges among the graphs. It adjusts weights *only* on edges that differ between the graphs, such that the queries' costs, when recomputed, will satisfy the ordering constraints provided by feedback.

## 5. EXPERIENCES AND OPEN PROBLEMS

The major innovation of CopyCat and the smart copy and paste model is its unified, lightweight interface for performing a range of information integration tasks — each of which has typically been addressed by a separate tool (often with its own learning component). In a sense, smart copy and paste is to data integration what a spreadsheet is to the database: a dynamic, user-editable workspace where data can be rapidly added, visualized, and reorganized.

CopyCat, at its heart, is a framework for plugging in information extractors, source description learners, and query learners under a common seamless interface. As described earlier in this paper, we were largely able to plug in state-of-the-art components developed previously by our research groups. Such components have been experimentally validated in isolation: for instance, the information extraction components are a major element of Fetch Technologies' business; query auto-completions (as implemented in the Karma system [36]) saved approximately 75% of keystrokes compared to manual integration of data by copy and paste; and learning of correct queries based on user feedback over answers converges very quickly in real domains such as biology [34] (as little as one item of feedback for a single query, and feedback on 10 queries to learn rankings for an entire family of queries).

Not surprisingly, the primary challenges in developing CopyCat were related to the user interface and experience, and the integrated processing of feedback across multiple learners and modules. Wherever possible, our goal was to follow the spreadsheet metaphor: the user should be able to (1) ignore auto-complete suggestions and continue to paste new data, (2) directly modify attribute labels or even modify imported data, or (3) provide direct feedback.

We believe the current version of CopyCat is really just the first step in a series of developments on the smart copy and paste model. Our initial experiences suggest many avenues of future work specifically on the SCP model — in addition to all of the challenges posed by the individual learning components (which are already the subject of work by the database and machine learning communities). We briefly discuss what we believe are the most important directions of work on SCP.

**Increased complexity and scale.** The types of integration scenarios we target with SCP are generally "lightweight" tasks that may involve a limited number of tasks and relatively small sources. As we increase the number of sources, there will be increasingly many possible queries and extractors. Open questions are how to present this to the user, such that it remains manageable and understandable, and how to ensure that there is sufficient information for the learner to make useful decisions.

**Advanced interactions.** To make the learning problem more tractable and to make interactions less complex, CopyCat makes certain default assumptions (e.g., that a set of sources should be joined using the conjunction of all possible predicates). In some cases an advanced user might want to remove some of these assumptions. In general we must consider how to balance between simplifying user choice and "overcommitting" to certain types of queries. Moreover, it will ultimately be important to allow advanced users to "undo" or edit certain portions of what they have demonstrated to the system and, perhaps, even to see how each demonstration changes the set of auto-completions.

**Complex functions / transforms.** Sometimes the user will want to apply complex operations that are difficult to demonstrate: for instance, perform an aggregation or evaluate an arithmetic expression. It is important to explore approaches to searching for possible functions [19], and also potentially to allow an advanced user to directly input such functions as in a spreadsheet.

**Feedback interaction.** Our user interface currently sends feedback to specific learners depending on whether we are in the "import" or "integrate" mode. We believe that ultimately there should be mechanisms for the integration learner to pass feedback from the integration mode to the source learners, and vice versa. To the best of our knowledge, little research has been done on enabling learners to cooperate.

**Data cleaning.** Our current implementation of CopyCat focuses on tasks relating to integrating data, but it does not include capabilities for editing the data once integrated — i.e., it does not support manual data cleaning. The Karma system [36] showed that our basic interface could be expanded to include data editing and cleaning capabilities. In CopyCat, the user would need to explicitly tell the system to switch into "cleaning" mode, so the system does not try to generalize any updates beyond the current tuple. An open question is whether the system can automatically determine when the user is cleaning a single tuple, versus making changes that should be generalized.

## 6. RELATED WORK

Best-effort information integration is sometimes considered the "next frontier" for information management: the term *dataspaces* [11] was proposed as a name for this vein of work. Research in this area has included support for lightweight and community-driven extraction [32] and mapping [18, 27], hints and trails [30], probabilistic schema mappings [10] and mediated schemas [31], and various efforts to integrate keyword search with integration. An excellent tutorial appeared in VLDB 2008 [15]. The idea of enabling non-expert users to create mashups has also been explored in the Potluck project [16], in this case using drag-and-drop instead of learning. The problem of learning extractors or wrappers for data sources has been heavily studied in both the machine learning and database communities, for example [2, 3, 12, 21, 26, 28]. Showing data provenance to assist in debugging and diagnosis was first proposed in [6], although to the best of our knowledge we are the first to make use of provenance to help learn from user feedback.

Our focus is on developing an *integrated creation and query system with provenance and feedback* — abolishing the divide between design-time, runtime, and debugging stages. This goal of integrated processing in CopyCat required significant extensions of the ideas and techniques first developed as stand-alone components in the authors' previous work: programming-by-demonstration techniques from the Karma [36, 35] mashup construction system, learning from feedback in the Q [34] ranked query answering system, learning techniques for information extraction [12, 28], and learning of source models [5, 24].

Our data integration framework incorporates a combination of programming by demonstration [9, 22, 25] and query by example (QBE) [37], and it is based on our earlier work on building mashups by example in Karma. In programming by demonstration, methods and procedures are induced from users' examples and interaction. This approach can be effective in various domains [13, 23, 33] where users understand and know how to do such tasks. In CopyCat users may not know how to formulate queries and only interact with the system through the data. The interaction is in a table similar to previous work on QBE. However, QBE requires users to manually select data sources, while CopyCat induces the sources to use by example and guides users to fill in only valid values.

## 7. CONCLUSIONS

This paper has described a new, unified model of information integration that removes the separation between design-time and runtime tasks and components. It is based on a model of monitoring copy operations from applications, suggesting generalizations and auto-completions through machine learning, and processing feedback. The smart copy and paste model is especially appropriate for lightweight integration tasks, where the integrator can immediately see the effects of each design decision, and can refine accordingly. We have identified a number of key research directions to be pursued. Our initial prototype, CopyCat, validates the basic framework and user interface, and provides a strong foundation upon which further work can be developed.

## 8. APPENDIX: CIDR DEMONSTRATION

At the conference we will demonstrate a working prototype of CopyCat in a way that highlights the user interaction and the learning components. In keeping with the examples of this paper, the data integration domain will be emergency response. The goals will be to demonstrate the following capabilities.

**Learning extractors and source descriptions** over both Web and source data. We will show how a few user-pasted examples are gen-eralized into extraction rules, for both spreadsheet and Web data. We will show how types are learned and inferred.

**Providing auto-completion results** for queries, based on user examples and known data sources, including joins, unions, and unit conversion. We will show how how the integration learner chooses among different potential associations, both among the data sources and among built-in services, in order to generate potential auto-complete queries.

**Providing visual explanations** for tuples, by taking the provenance of each query and mapping it into an intuitive graphical representation. This representation captures "alternative explanations" (when a tuple is produced by more than one query) as well as explanations involving multiple joins.

**Exporting data** to common application formats, including XML and, perhaps more interestingly, the Google Maps interface. This capability makes it very easy to use CopyCat as a mashup generator.

We now briefly summarize the data sources and the task we will use to illustrate these capabilities. Our goal is to use live data from the Web if an Internet connection is available, but we have saved local HTML pages.

## 8.1 Data Sources

Our data sources will include real Web pages with shelter information, such as the one in Figure 1; Excel spreadsheets with contact information for the shelters; and address resolution and geocoding services from the Web (e.g., Google or Yahoo).

## 8.2 Task

We will perform an integration task where the CopyCat system knows nothing about the data sources, and hence has no extractors for them. The goal will be to plot shelters on a map, such that a user might be able to determine which shelters are accessible from his or her area. This is an example of an integration task specifically driven by a query, rather than a task to create a mediated schema.

In keeping with the spirit of smart copy and paste, our task will be achieved simply by copying and pasting data from the sources — as opposed to invoking a series of disparate tools (e.g., wrapper induction or entering complex pattern matching expressions, schema matching, schema merging, mapping construction, query processing).

In the process, we will show how the system infers types and names for pasted attributes; makes suggestions for useful attributes to be added to the integration query; auto-completes rows by extracting additional data from sources; and finds potential integration queries. We will demonstrate the query explanation facility for different rows, and show how feedback is processed to remove irrelevant answers and commit relevant ones. We will demonstrate explicit feedback on auto-complete suggestions, as well feedback based on ignoring the existing suggestions and simply pasting or correcting data.

## 8.3 Summary

This demonstration should provide a good sense of the user interaction and capabilities of the CopyCat system across a variety of data sources, and to show the utility of both provenance and feedback in the system. The live demonstration should add significant value above and beyond the talk, as it is precisely the user interaction mode that is the focus of smart copy and paste.

## ACKNOWLEDGMENTS

The research in this paper has been funded in part by NSF grants IIS-0477972, 0513778, and 0415810, in part by the DARPA DIESEL effort, and in part by DARPA under Contract No. FA8750-07-D-0815/0004. We thank Koby Crammer, Fernando Pereira, Sudipto Guha, and M. Salman Mehmood for their contributions to the query learning components used within the integration learner of this paper.

## 9. REFERENCES

[1] J. L. Ambite, C. A. Knoblock, K. Lerman, A. Plangprasopchok, T. Russ, C. Gazen, S. Minton, and M. Carman. Exploiting data semantics to discover, extract, and model web sources. In *Proceedings of the First International Workshop on Semantic Aspects in Data Mining (SADM'08)*, 2008.

[2] N. Ashish and C. A. Knoblock. Semi-automatic wrapper generation for Internet information sources. In *Proceedings of the Second IFCIS International Conference on Cooperative Information Systems, Kiawah Island, South Carolina, USA, June 24-27, 1997*, 1997.

[3] R. Baumgartner, S. Flesca, and G. Gottlob. Visual web information extraction with Lixto. In *VLDB*, 2001.

[4] P. Buneman, S. B. Davidson, M. F. Fernandez, and D. Suciu. Adding structure to unstructured data. In *ICDT*, volume 1186, 1997.

[5] M. Carman and C. A. Knoblock. Learning semantic definitions of online information sources. *Journal of Artificial Intelligence Research (JAIR)*, 30, 2007.

[6] L. Chiticariu and W.-C. Tan. Debugging schema mappings with routes. In *VLDB*, 2006.

[7] K. Crammer, O. Dekel, J. Keshet, S. Shalev-Shwartz, and Y. Singer. Online passive-aggressive algorithms. *Journal of Machine Learning Research*, 7:551–585, 2006.

[8] Y. Cui. *Lineage Tracing in Data Warehouses*. PhD thesis, Stanford University, 2001.

[9] A. Cypher, D. C. Halbert, D. Kurlander, H. Lieberman, D. Maulsby, B. A. Myers, and A. Turransky, editors. *Watch What I Do: Programming by Demonstration*. The MIT Press, 1993. Available from http://acypher.com/wwid/.

[10] X. L. Dong, A. Y. Halevy, and C. Yu. Data integration with uncertainty. In *VLDB*, 2007.

[11] M. Franklin, A. Halevy, and D. Maier. From databases to dataspaces: a new abstraction for information management. *SIGMOD Rec.*, 34(4), 2005.

[12] B. Gazen and S. Minton. Overview of autofeed: An unsupervised learning system for generating webfeeds. In *AAAI*, 2006.

[13] A. Gibson, M. Gamble, K. Wolstencroft, T. Oinn, and C. Goble. The data playground: An intuitive workflow specification environment. In *E-SCIENCE*, 2007.

[14] A. Y. Halevy, Z. G. Ives, D. Suciu, and I. Tatarinov. Schema mediation in peer data management systems. In *ICDE*, March 2003.

[15] A. Y. Halevy, D. Maier, and M. J. Franklin. Dataspaces: The tutorial. In *VLDB*, 2008.

[16] D. F. Huynh, R. C. Miller, and D. R. Karger. Potluck: Data mash-up tool for casual users. In *ISWC/ASWC*, 2007.

[17] Z. G. Ives, T. J. Green, G. Karvounarakis, N. E. Taylor, V. Tannen, P. P. Talukdar, M. Jacob, and F. Pereira. The ORCHESTRA collaborative data sharing system. *SIGMOD*

*Rec.*, 2008.

[18] S. R. Jeffery, M. J. Franklin, and A. Y. Halevy. Pay-as-you-go user feedback for dataspace systems. In *SIGMOD*, 2008.

[19] H. Kache, Y. Saillet, and M. Roth. Transformation rule discovery through data mining. In *International Workshop on New Trends in Information Integration*, 2008.

[20] V. Kacholia, S. Pandit, S. Chakrabarti, S. Sudarshan, R. Desai, and H. Karambelkar. Bidirectional expansion for keyword search on graph databases. In *VLDB*, 2005.

[21] N. Kushmerick, R. Doorenbos, and D. Weld. Wrapper induction for information extraction. In *IJCAI '97*, 1997.

[22] T. Lau. *Programming by Demonstration: a Machine Learning Approach*. PhD thesis, University of Washington, 2001.

[23] T. Lau, L. Bergman, V. Castelli, and D. Oblinger. Sheepdog: learning procedures for technical support. In *IUI*, 2004.

[24] K. Lerman, A. Plangprasopchok, and C. A. Knoblock. Semantic labeling of online information sources. *International Journal on Semantic Web and Information Systems*, 3(3), 2007.

[25] H. Lieberman, editor. *Your Wish is My Command: Programming by Example*. Morgan Kaufman, 2001.

[26] L. Liu, C. Pu, and W. Han. XWRAP: An XML-enabled wrapper construction system for web information sources. In *ICDE*, 2000.

[27] R. McCann, W. Shen, and A. Doan. Matching schemas in online communities: A web 2.0 approach. In *ICDE*, 2008.

[28] I. Muslea, S. Minton, and C. A. Knoblock. Hierarchical wrapper induction for semistructured information sources. *Autonomous Agents and Multi-Agent Systems*, 4(1/2), 2001.

[29] E. Rahm and P. A. Bernstein. A survey of approaches to automatic schema matching. *VLDB J.*, 10(4), 2001.

[30] M. A. V. Salles, J.-P. Dittrich, S. K. Karakashian, O. R. Girard, and L. Blunschi. iTrails: pay-as-you-go information integration in dataspaces. In *VLDB*, 2007.

[31] A. D. Sarma, X. Dong, and A. Halevy. Bootstrapping pay-as-you-go data integration systems. In *SIGMOD*, New York, NY, USA, 2008.

[32] W. Shen, P. DeRose, R. McCann, A. Doan, and R. Ramakrishnan. Toward best-effort information extraction. In *SIGMOD*, New York, NY, USA, 2008.

[33] A. Sugiura and Y. Koseki. Internet scrapbook: automating web browsing tasks by demonstration. In *UIST*, 1998.

[34] P. P. Talukdar, M. Jacob, M. S. Mehmood, K. Crammer, Z. G. Ives, F. Pereira, and S. Guha. Learning to create data-integrating queries. In *VLDB*, 2008.

[35] R. Tuchinda, P. Szekely, and C. A. Knoblock. Building data integration queries by demonstration. In *IUI*, 2007.

[36] R. Tuchinda, P. Szekely, and C. A. Knoblock. Building mashups by example. In *IUI*, 2008.

[37] M. M. Zloof. Query by example: A data base language. *IBM Systems Journal*, 16(4), 1977.